%% file: ms.tex
\documentclass[12pt,preprint]{aastex}

\newcommand{\teff}{$T_{\rm eff}$}

\shorttitle{Abundances in NGC 1851}
\shortauthors{Yong \& Grundahl}

\begin{document}

\title{An abundance analysis of bright giants in the globular 
cluster NGC 1851\footnote{Based on observations made with ESO Telescopes 
at the Paranal Observatories under programme ID 70.B-0361(A)}}

\author{David Yong}
\affil{Research School of Astronomy and Astrophysics, Australian National 
University, Weston, ACT 2611, 
Australia}
\email{yong@mso.anu.edu.au}

\and

\author{Frank Grundahl}
\affil{Department of Physics and Astronomy, University of Aarhus, Denmark}
\email{fgj@phys.au.dk}

\begin{abstract}

We present the chemical compositions for eight bright giants in the globular 
cluster NGC 1851. Our analysis reveals large star-to-star abundance 
variations and correlations of the light elements O, Na, and Al, 
a feature found in every well 
studied globular cluster. However, NGC 1851 also exhibits 
large star-to-star abundance variations of the $s$-process elements Zr 
and La. These $s$-process elements are correlated with Al, and anticorrelated 
with O. Furthermore, the Zr and La abundances appear to cluster around 
two distinct values. 
A recent study revealed a double subgiant branch in NGC 1851. 
Our data reinforce the notion that there are two stellar 
populations in NGC 1851 
and indicate that this cluster has experienced 
a complicated formation history with similarities 
to $\omega$ Centauri. 

\end{abstract}


\section{Introduction}
\label{sec:intro}

Recent studies have revealed that the Galactic globular clusters 
$\omega$ Centauri \citep{bedin04}, 
NGC 2808 \citep{dantona05,piotto07}, and M 54 \citep{siegel07}
contain multiple stellar populations. The evidence consists of 
HST color-magnitude diagrams (CMDs) in which multiple main sequences have 
been identified. 
In all cases, the most plausible explanation is that these globular clusters 
contain populations with distinct compositions and/or ages, in 
contrast to the classical concept that globular clusters are 
monometallic, coeval, simple stellar 
populations. The most recent addition to this intriguing collection 
is NGC 1851, a globular cluster with a bimodal horizontal branch (HB),  
whose CMD displays a double subgiant branch \citep{milone07}. 

Both $\omega$ Centauri and M 54 possess a large metallicity spread and 
are the two most massive Galactic globular clusters. 
$\omega$ Centauri is widely regarded as the nucleus of an accreted dwarf 
galaxy (e.g., \citealt{smith00}) and 
M 54 is suspected 
to be the nucleus of the Sagittarius dwarf spheroidal galaxy 
(e.g., \citealt{layden00}). However, not all clusters with 
multiple main sequences show a metallicity spread and although there is 
speculation that all globular clusters may have been the nuclei of 
ancient dwarf galaxies \citep{bekki06}, currently there is 
little direct observational evidence. 
For NGC 2808, the multiple main sequences and complex HB morphology
can be attributed to a large He abundance variation 
\citep{dantona04,dantona05,piotto07}, 
but the Fe abundances and ages show little, if any, 
variation among the populations. Although 
there are large star-to-star 
abundance variations of 
O, Na, and Al in NGC 2808 \citep{carretta06}, such 
patterns are ubiquitous in Galactic globular clusters (e.g., 
\citealt{kraft94} and \citealt{gratton04}). 

The situation regarding the chemical compositions for stars in the
globular cluster NGC 1851 
is less clear due to the lack of data. The sole spectroscopic 
abundance analysis was conducted by \citet{hesser82}. Low resolution 
spectra revealed extremely strong CN bands for three out of eight 
bright red giants. \citet{hesser82} also found that the Sr and Ba 
lines may be enhanced in CN-strong stars relative to 
other giants in this cluster. Such abundance patterns, if confirmed, 
would bear a striking resemblance to $\omega$ Centauri. 
A study of the stellar chemical compositions in NGC 1851 
is of great interest in order 
to provide constraints upon the interpretation of the double 
subgiant branch and to explore the chemical evolution of this cluster.

\section{Observations and abundance analysis}
\label{sec:data}

The spectra for eight bright giants in NGC 1851 
were retrieved from the ESO archive. The observations 
were obtained using UVES \citep{uves}. Each star was observed for 
900 seconds using the 0.70\arcsec~slit which provided 
a spectral resolution of R $\simeq$ 55,000 per 4 pixel resolution element. 
The data were reduced using standard procedures in IRAF. The 
wavelength coverage ranged from 5950\AA~to 9750\AA~with a 
typical signal-to-noise ratio of 60 per pixel near 6500\AA. All stars 
are radial velocity members with $v_r$ ranging from +316 km/s to +332 km/s. 
One star, 333, was also observed by \citet{hesser82}.  

The stellar parameters were determined using the same tools and techniques
described in \citet{grundahl02} and \citet{yong06}. 
While the agreement between 
the photometric and spectroscopic stellar parameters was excellent, we 
adopted the spectroscopic parameters. (Photometric estimates were based on 
data from \citet{grundahl99} and 2MASS \citep{2mass}. 
The results and conclusions 
would not change had we adopted photometric stellar parameters.) 

Abundances 
were measured from an equivalent width analysis. For O, La, and Eu, 
synthetic spectra were generated to account for blends, 
hyperfine structure, and/or isotopic splitting. We used 
the LTE line analysis and spectrum synthesis program MOOG \citep{moog}, 
\citet{kurucz93} model atmospheres, and the 
same line lists employed by \citet{yong05}. 
While many lines of interest 
lie below 6000\AA~and were not observed, there were a sufficient 
number of lines from which the abundances of O, Na, Mg, Al, Si, Ca, Sc, Ti, 
V, Mn, Fe, Co, Ni, Zr, La, and Eu could be measured. 
We focus upon the most fascinating results which come 
from the light elements (O, Na, and Al), Fe, and the neutron-capture 
elements (Zr, La, and Eu). The stellar parameters and abundances are 
presented in Table \ref{tab:abund}. 

\section{Results }
\label{sec:results}

\begin{figure}
\epsscale{0.8}
\plotone{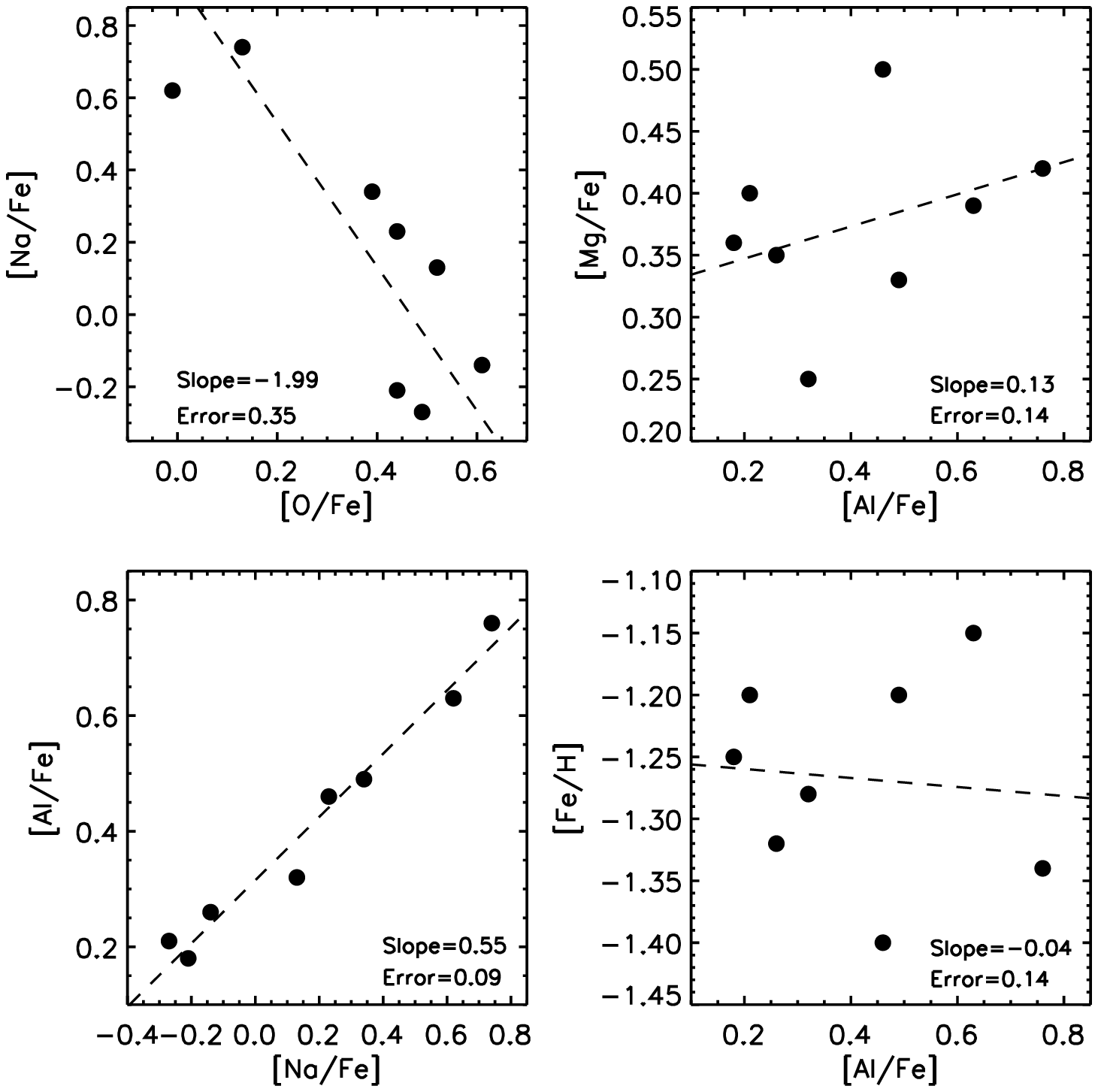}
\caption{Clockwise from upper left, [Na/Fe] vs.\ [O/Fe], 
[Mg/Fe] vs.\ [Al/Fe], [Fe/H] vs.\ [Al/Fe], and 
[Al/Fe] vs.\ [Na/Fe]. Linear least squares fits to the data 
are shown (slope and associated error are included). 
O and Na are anticorrelated and 
Al and Na are correlated. Mg and Al do not appear to be 
correlated nor are Fe and Al.\label{fig:onamgalfe}}
\end{figure}

Even within our small sample of stars, NGC 1851 exhibits large 
star-to-star abundance 
variations of the light elements O, Na, and Al 
(see Figure \ref{fig:onamgalfe}). 
The amplitude of the variation is 0.62, 1.02, and 0.58 dex for 
[O/Fe], [Na/Fe], and [Al/Fe] respectively. 
O and Na are anticorrelated, Na and Al are 
correlated, and there does not appear to be an 
anticorrelation between Mg and Al. 
We note that the correlations remain evident regardless of 
whether we plot [X/Fe] or log $\epsilon$(X). 
Such abundance patterns have been found in all globular 
clusters, including those with and without multiple main sequences. 

We find a mean abundance 
[Fe/H] = $-$1.27 $\pm$ 0.03 ($\sigma$ = 0.09) with values spanning 0.25 
dex. 
The ratio [Fe/H] appears to exhibit a range that may be 
larger than that expected from observational uncertainties. 
It is not 
clear whether this is a real (though small) 
spread in the Fe abundance, or merely an artefact of small number statistics
or the limited set of Fe lines available 
given the restricted wavelength coverage. 
Figure \ref{fig:onamgalfe} 
shows that the Fe abundance is not correlated with the Al abundance. 

\begin{figure}
\epsscale{0.8}
\plotone{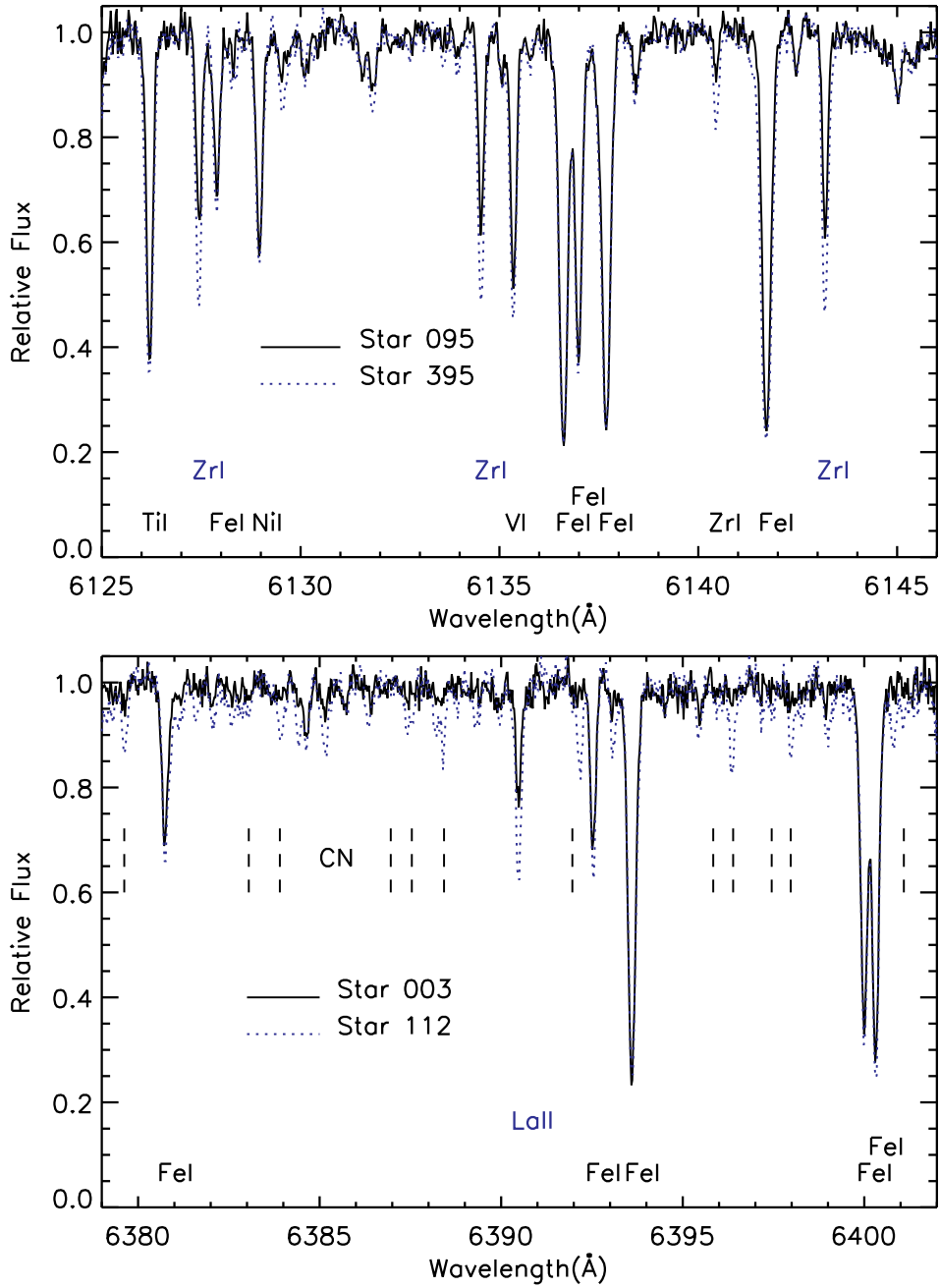}
\caption{The upper panel shows stars 095 and 395 centered on the 
Zr\,{\sc ii} lines. The stellar parameters are very similar as 
confirmed by the atomic lines of similar strength. However, 
the Zr lines differ considerably. 
The lower panel shows stars 003 and 112 
centered on the La\,{\sc ii} line. The stellar parameters are very 
similar as are the strengths of atomic lines. However, the La line 
differs considerably as do the CN molecular lines. 
\label{fig:zrla}}
\end{figure}

The most striking result of our analysis is that NGC 1851 has 
a significant star-to-star abundance variation of the light $s$-process 
element Zr and the heavy $s$-process element La.  
Figure \ref{fig:zrla} shows two pairs 
of stars with essentially identical stellar parameters. Therefore, 
any difference in line strengths must be due to differences in 
abundances between the pairs of stars. In this Figure, the 
strengths of Fe and other atomic lines are very similar between the pairs of 
stars, as expected. However, the Zr and La lines show large differences and 
therefore the Zr and La abundances must differ considerably. 
From a quantitative analysis, the amplitude of the 
variation is 0.56 and 0.49 dex for [Zr/Fe] and [La/Fe] respectively. 
This Figure also shows that star 112 is CN-strong (and La-rich) relative 
to star 003. Recall that \citet{hesser82} found extremely CN-strong 
stars in this cluster and that the Sr and Ba lines were enhanced
in these stars. 
Unfortunately Zr and La are the only $s$-process elements 
that we could measure due to the wavelength 
coverage and we were also unable to measure C or N 
abundances. We note that the strengths of the 6141.71\AA~and 
6496.90\AA~Ba\,{\sc ii} lines differ from star-to-star with the Zr- and 
La-rich stars exhibiting stronger Ba lines. However, the Ba lines 
are saturated ($>$ 190m\AA) such that reliable abundances cannot be 
readily 
measured and therefore we cannot quantify the star-to-star 
Ba abundance dispersion. According to \citet{hesser82}, star 333 
is CN-normal and we find no enhancement of Zr or La. 

\begin{figure}
\epsscale{0.8}
\plotone{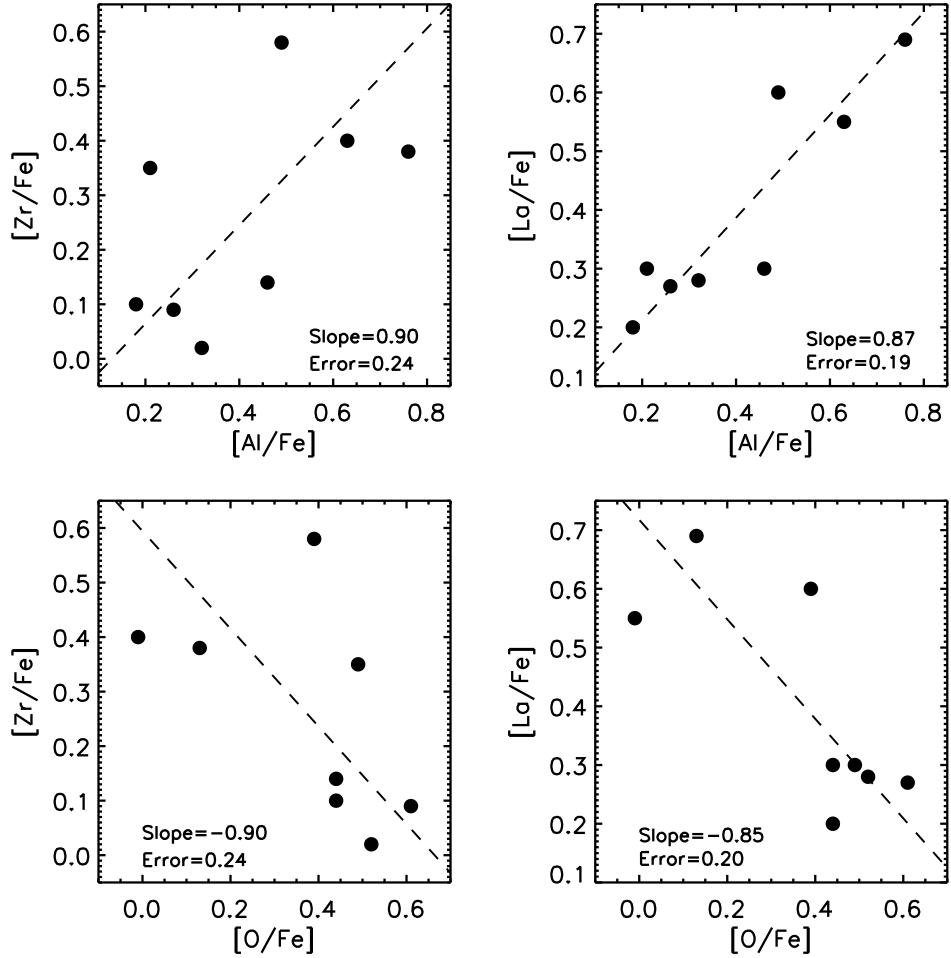}
\caption{Clockwise from upper left, [Zr/Fe] vs.\ [Al/Fe], 
[La/Fe] vs.\ [Al/Fe], [La/Fe] vs.\ [O/Fe], and 
[Zr/Fe] vs.\ [O/Fe]. Linear least squares fits to the data 
are shown (slope and associated error are included).
Zr and La increase 
with increasing Al and decrease with increasing 
O.\label{fig:oalzrla}}
\end{figure}

In Figure \ref{fig:oalzrla}, the $s$-process elements Zr and La are 
correlated with Al and anticorrelated with O. Once again, 
the correlations are present regardless of whether we plot [X/Fe] or 
log $\epsilon$(X). 
The abundances of 
Zr and La cluster around two distinct values. The three La-rich stars
have a mean value [La/Fe] = 0.61 while the five La-normal stars have 
a mean value [La/Fe] = 0.27. The three La-rich stars are also Zr-rich 
with a mean value [Zr/Fe] = 0.45 and the five La-normal stars are also 
Zr-normal with a mean value [Zr/Fe] = 0.14. 
Curiously, one of the La-normal stars (star 209)
appears to be relatively Zr-rich, [Zr/Fe] = 0.35. 
Setting aside this star, the four remaining 
Zr-normal stars have a mean value [Zr/Fe] = 0.09. 
The mean ratio of heavy to light $s$-process elements is very similar 
for the three La-rich, Zr-rich stars and for the
five La-normal, Zr-normal stars, [La/Zr] = 0.16 and 0.13 respectively. 

Finally, we draw attention to the abundance of 
the $r$-process element Eu. The average ratio,
[Eu/Fe] = +0.71, is rather high compared to other globular 
clusters and field stars at the metallicity of NGC 1851 
(e.g., \citealt{pritzl05}). We do not find any evidence for a 
star-to-star Eu abundance dispersion or bimodality. Eu 
is not correlated with the light elements O or Al nor with the 
$s$-process elements Zr and La. 

\section{Discussion}
\label{sec:discussion}

\citet{milone07} explored various possibilities for explaining 
the double subgiant branch in NGC 1851. 
If age is the sole parameter, a difference of 1 Gyr would be required. 
If composition is the sole parameter, any abundance variation invoked 
to account for the double subgiant branch 
is heavily constrained by the width of the main sequence (and giant branch). 
\citet{milone07} estimate that the maximum 
possible abundance variation on the main sequence would be 
$\Delta$[Fe/H] = 0.1 dex or a helium abundance $\Delta Y$ = 0.026. 
They argue that the He abundance alone cannot explain the 
observed subgiant branch. They suggest that a population with a 0.2 dex 
metallicity increase could explain the subgiant branch, but that such 
an abundance spread is precluded by the width of the main sequence and 
giant branch. However, a combination of increased 
[Fe/H] by 0.2 dex and helium from Y = 0.247 to Y = 0.30 could 
reproduce the observed 
subgiant branch behavior as well as ensuring a sufficiently 
narrow main sequence and giant branch. 

Available measurements indicate that the abundances of Fe, O, Na, Al, and 
neutron capture elements 
do not change as a function of evolutionary status from the main sequence 
to the giant branch in globular clusters (e.g., \citealt{gratton01}, 
\citealt{james04} and 
\citealt{cohen05}). If we assume that the abundances for giant stars 
are representative of subgiant stars in NGC 1851, then our measured
abundances eliminate the 
possibility that the subgiant branch split is due to an abundance 
difference of 0.2 dex for Fe and a shift of $Y$ from 0.247 to 0.30. 
While our sample size is limited, there is no hint of a bimodal 
distribution of Fe abundances separated by 0.2 dex, as required for the 
chemical composition explanation 
for the double subgiant branch. Indeed, \citet{milone07} noted 
that such a possibility is also precluded by the 
magnitudes of blue HB stars. Interestingly, the dispersion in 
our Fe abundances ($\sigma$ = 0.09 dex) is within the 0.1 dex 
``limit'' imposed by the width of the main sequence. 

The abundances of additional elements 
add further complexity to NGC 1851. The star-to-star 
abundance variation of O, Na, and Al and the anticorrelation of O and Na 
exist in NGC 1851 and indeed in all globular clusters. While our sample size 
is small, the amplitude of these abundance variations 
is comparable to clusters at a 
similar metallicity. However, the light $s$-process element 
Zr and the heavy $s$-process element La exhibit a large star-to-star 
abundance variation. (Our spectra also indicate a large range in 
Ba line strengths, and presumably Ba abundance, but we 
are unable to quantify the range.) 
Furthermore, the Zr and La abundances appear 
to have a bimodal distribution and reinforce the idea that there 
are two stellar populations in this cluster. (The ratio of 
heavy to light $s$-elements, [La/Zr] is similar for the 
``two populations''.) 
La and Al (and Zr and Al) are correlated and so the source of the 
Al variation is likely to be the source of the La 
variation. The hint of a correlation between Al and Zr has also been found 
in NGC 6752 \citep{yong05}. 
Asymptotic giant branch (AGB) stars may produce large 
enhancements of Al and La, albeit via different processes in 
AGB stars of different masses (e.g., \citealt{busso99}, \citealt{karakas03}). 
The He, C, N, and O 
produced by such AGB stars may play a role on the age differences 
for the subgiant branch populations. 

\citet{milone07} raise the tantalizing prospect that the relative 
frequency of stars on the fainter/brighter subgiant branches (45\% 
vs.\ 55\%) roughly matches the relative frequency of HB stars 
bluer/redder than the instability strip (37\% vs.\ 63\%). The fraction of 
CN-strong, Ba-strong, Sr-strong stars found by \citet{hesser82} was 
$\sim$ 40\%. 
In this study, the fraction of Zr-strong, La-strong stars was also 
$\sim$ 40\%. Based on the relative numbers, we speculate that the 
brighter subgiant branch stars have ``normal'' CN, Sr, Zr, Ba, and La
abundances and populate the red HB. We also speculate that the 
fainter subgiant branch stars are CN-strong, enriched in Sr, Zr, Ba, 
and La, and populate the blue HB. Indeed if the CN, Zr, and La enriched 
stars are also He-rich, they are expected to populate the 
blue HB \citep{sweigart97}. However, we caution that the relative 
numbers of the two populations will not be the same at all 
evolutionary phases. Nevertheless, the abundances in  
subgiant branch, giant, and red HB stars can test this hypothesis. 

Since the fainter subgiant branch stars are older than the brighter subgiant 
branch stars (assuming age is the only parameter controlling the double 
subgiant branch), then our tentative speculation would require that the 
Zr-rich, La-rich stars are younger than the Zr-normal, La-normal 
stars. While this unpalatable scenario would be a challenge to 
explain, \citet{villanova07} find that the most metal-rich subgiant 
stars in $\omega$ Centauri are among the oldest stars in the cluster. 
Clearly, the sequence of events that led to the formation of the 
different stellar populations in NGC 1851 and $\omega$ Centauri 
requires some imagination. 
On the other hand, an analysis of a larger sample of giant stars may reveal 
that the CN-strong, Zr-rich, La-rich stars are associated with the 
majority population of brighter subgiant branch stars. In this case, 
if the CN-strong, Zr-rich, La-rich stars are the progeny of the 
CN-normal, Zr-normal, La-normal stars, 
the assumed 1 Gyr age difference would then place constraints upon the 
mass range of the AGB stars that produced the CN, Zr, and La 
excess. At the metallicity of NGC 1851, 
Karakas (2007 priv.\ comm.) suggests that the minimum mass (ZAMS) 
of the AGB stars would be approximately 2 M$_{\odot}$. Further, the 
AGB stars that produce La have M $\gtrsim$ 1 M$_\odot$ and the 
AGB stars that produce Al have M $\gtrsim$ 4 M$_\odot$. 

Finally, we note that the only other globular cluster 
that exhibits a large range in C, N, O, Na, Al, Sr, Zr, Ba, and La 
abundances is $\omega$ Centauri. Both NGC 1851 and $\omega$ Centauri
display multiple subgiant branches which are presumably due to 
stellar populations with different ages and compositions. 
Further, NGC 1851 and $\omega$ Centauri are the only clusters 
that display a large variation in the 
Str{\" om}gren $m_1$ index, traditionally used as a metallicity
indicator. NGC 1851 appears to exhibit a bimodal $m_1$ distribution 
on the giant branch suggesting an extreme variation in C and/or N 
abundances (Grundahl 2007 in preparation). 
In the context of the formation of the Galactic globular cluster 
system, we speculate that NGC 1851 may represent a 
``bridge'' between $\omega$ Centauri 
(large variation of all elements) and NGC 6752-like clusters 
(constant Fe but large variations of light elements C-Al). 
A more detailed analysis of additional elements and additional stars 
in NGC 1851 is of great interest. 

\acknowledgments

This research has made use of the SIMBAD database,
operated at CDS, Strasbourg, France and
NASA's Astrophysics Data System. We thank
Gary Da Costa, Amanda Karakas, John Norris, and the anonymous referee 
for helpful comments. 
FG gratefully acknowledges financial support from 
The Danish AsteroSeismology Centre (DASC), Carlsbergfondet, 
Instrumentcenter for Dansk Astronomi (IDA), and 
the project ``Stars: Central engines of the evolution of the Universe'',
carried out at Aarhus University and Copenhagen University, supported
by the Danish National Science Research Council.
This research was 
supported in part by NASA through the American Astronomical Society's Small 
Research Grant Program.

\clearpage

\input{tab1}

\clearpage

\clearpage

\clearpage

\end{document}

%% file: tab1.tex
\begin{deluxetable}{cccccrrccccc}
\rotate
\tablecolumns{12} 
\tablewidth{0pc} 
\tablecaption{Stellar parameters and abundances\label{tab:abund}} 
\tablehead{ 
\colhead{Star\tablenotemark{a}} & 
\colhead{\teff~(K)} & 
\colhead{$\log g$ (cgs)} & 
\colhead{$v_t$ (km s$^{-1}$)} & 
\colhead{[Fe/H]} & 
\colhead{[O/Fe]} & 
\colhead{[Na/Fe]} & 
\colhead{[Mg/Fe]} & 
\colhead{[Al/Fe]} & 
\colhead{[Zr/Fe]} & 
\colhead{[La/Fe]} & 
\colhead{[Eu/Fe]}
}
\startdata 
003 & 4075 & 0.60 & 1.80 & $-$1.28 & 0.52 & 0.13 & 0.25 & 0.32 & 0.02 & 0.28 & 0.83 \\
095 & 4025 & 0.15 & 1.95 & $-$1.40 & 0.44 & 0.23 & 0.50 & 0.46 & 0.14 & 0.30 & 0.75 \\
112 & 4025 & 0.25 & 1.85 & $-$1.34 & 0.13 & 0.74 & 0.42 & 0.76 & 0.38 & 0.69 & 0.69 \\
151 & 4175 & 0.65 & 1.80 & $-$1.32 & 0.61 & $-$0.14 & 0.35 & 0.26 & 0.09 & 0.27 & 0.82 \\
209 & 4400 & 1.15 & 1.75 & $-$1.20 & 0.49 & $-$0.27 & 0.40 & 0.21 & 0.35 & 0.30 & 0.70 \\
329 & 3875 & 0.10 & 1.55 & $-$1.20 & 0.39 & 0.34 & 0.33 & 0.49 & 0.58 & 0.60 & 0.65 \\
333 & 4225 & 0.75 & 1.70 & $-$1.25 & 0.44 & $-$0.21 & 0.36 & 0.18 & 0.10 & 0.20 & 0.65 \\
395 & 3975 & 0.45 & 1.65 & $-$1.15 & $-$0.01 & 0.62 & 0.39 & 0.63 & 0.40 & 0.55 & 0.60 \\
\enddata 
\tablenotetext{a}{Star names taken from \citet{stetson81}.}
\end{deluxetable}

%% file: ms.bbl
\begin{thebibliography}{29}
\expandafter\ifx\csname natexlab\endcsname\relax\def\natexlab#1{#1}\fi

\bibitem[{{Bedin} {et~al.}(2004){Bedin}, {Piotto}, {Anderson}, {Cassisi},
  {King}, {Momany}, \& {Carraro}}]{bedin04}
{Bedin}, L.~R., {Piotto}, G., {Anderson}, J., {Cassisi}, S., {King}, I.~R.,
  {Momany}, Y., \& {Carraro}, G. 2004, \apjl, 605, L125

\bibitem[{{Bekki}(2006)}]{bekki06}
{Bekki}, K. 2006, \mnras, 367, L24

\bibitem[{{Busso} {et~al.}(1999){Busso}, {Gallino}, \& {Wasserburg}}]{busso99}
{Busso}, M., {Gallino}, R., \& {Wasserburg}, G.~J. 1999, \araa, 37, 239

\bibitem[{{Carretta}(2006)}]{carretta06}
{Carretta}, E. 2006, \aj, 131, 1766

\bibitem[{{Cohen} \& {Mel{\'e}ndez}(2005)}]{cohen05}
{Cohen}, J.~G. \& {Mel{\'e}ndez}, J. 2005, \aj, 129, 303

\bibitem[{{D'Antona} {et~al.}(2005){D'Antona}, {Bellazzini}, {Caloi}, {Pecci},
  {Galleti}, \& {Rood}}]{dantona05}
{D'Antona}, F., {Bellazzini}, M., {Caloi}, V., {Pecci}, F.~F., {Galleti}, S.,
  \& {Rood}, R.~T. 2005, \apj, 631, 868

\bibitem[{{D'Antona} \& {Caloi}(2004)}]{dantona04}
{D'Antona}, F. \& {Caloi}, V. 2004, \apj, 611, 871

\bibitem[{{Dekker} {et~al.}(2000){Dekker}, {D'Odorico}, {Kaufer}, {Delabre}, \&
  {Kotzlowski}}]{uves}
{Dekker}, H., {D'Odorico}, S., {Kaufer}, A., {Delabre}, B., \& {Kotzlowski}, H.
  2000, in Presented at the Society of Photo-Optical Instrumentation Engineers
  (SPIE) Conference, Vol. 4008, Proc. SPIE Vol. 4008, p. 534-545, Optical and
  IR Telescope Instrumentation and Detectors, Masanori Iye; Alan F. Moorwood;
  Eds., ed. M.~{Iye} \& A.~F. {Moorwood}, 534--545

\bibitem[{{Gratton} {et~al.}(2004){Gratton}, {Sneden}, \&
  {Carretta}}]{gratton04}
{Gratton}, R., {Sneden}, C., \& {Carretta}, E. 2004, \araa, 42, 385

\bibitem[{{Gratton} {et~al.}(2001){Gratton}, {Bonifacio}, {Bragaglia},
  {Carretta}, {Castellani}, {Centurion}, {Chieffi}, {Claudi}, {Clementini},
  {D'Antona}, {Desidera}, {Fran{\c c}ois}, {Grundahl}, {Lucatello}, {Molaro},
  {Pasquini}, {Sneden}, {Spite}, \& {Straniero}}]{gratton01}
{Gratton}, R.~G., {Bonifacio}, P., {Bragaglia}, A., {Carretta}, E.,
  {Castellani}, V., {Centurion}, M., {Chieffi}, A., {Claudi}, R., {Clementini},
  G., {D'Antona}, F., {Desidera}, S., {Fran{\c c}ois}, P., {Grundahl}, F.,
  {Lucatello}, S., {Molaro}, P., {Pasquini}, L., {Sneden}, C., {Spite}, F., \&
  {Straniero}, O. 2001, \aap, 369, 87

\bibitem[{{Grundahl} {et~al.}(2002){Grundahl}, {Briley}, {Nissen}, \&
  {Feltzing}}]{grundahl02}
{Grundahl}, F., {Briley}, M., {Nissen}, P.~E., \& {Feltzing}, S. 2002, \aap,
  385, L14

\bibitem[{{Grundahl} {et~al.}(1999){Grundahl}, {Catelan}, {Landsman},
  {Stetson}, \& {Andersen}}]{grundahl99}
{Grundahl}, F., {Catelan}, M., {Landsman}, W.~B., {Stetson}, P.~B., \&
  {Andersen}, M.~I. 1999, \apj, 524, 242

\bibitem[{{Hesser} {et~al.}(1982){Hesser}, {Bell}, {Harris}, \&
  {Cannon}}]{hesser82}
{Hesser}, J.~E., {Bell}, R.~A., {Harris}, G.~L.~H., \& {Cannon}, R.~D. 1982,
  \aj, 87, 1470

\bibitem[{{James} {et~al.}(2004){James}, {Fran{\c c}ois}, {Bonifacio},
  {Carretta}, {Gratton}, \& {Spite}}]{james04}
{James}, G., {Fran{\c c}ois}, P., {Bonifacio}, P., {Carretta}, E., {Gratton},
  R.~G., \& {Spite}, F. 2004, \aap, 427, 825

\bibitem[{{Karakas} \& {Lattanzio}(2003)}]{karakas03}
{Karakas}, A.~I. \& {Lattanzio}, J.~C. 2003, Publications of the Astronomical
  Society of Australia, 20, 279

\bibitem[{{Kraft}(1994)}]{kraft94}
{Kraft}, R.~P. 1994, \pasp, 106, 553

\bibitem[{{Kurucz}(1993)}]{kurucz93}
{Kurucz}, R. 1993, ATLAS9 Stellar Atmosphere Programs and 2 km/s grid.~Kurucz
  CD-ROM No.~13.~ Cambridge, Mass.: Smithsonian Astrophysical Observatory,
  1993., 13

\bibitem[{{Layden} \& {Sarajedini}(2000)}]{layden00}
{Layden}, A.~C. \& {Sarajedini}, A. 2000, \aj, 119, 1760

\bibitem[{{Milone} {et~al.}(2007){Milone}, {Bedin}, {Piotto}, {Anderson},
  {King}, {Sarajedini}, {Dotter}, {Chaboyer}, {Marin-Franch}, {Majewski},
  {Aparicio}, {Hempel}, {Paust}, {Reid}, {Rosenberg}, \& {Siegel}}]{milone07}
{Milone}, A.~P., {Bedin}, L.~R., {Piotto}, G., {Anderson}, J., {King}, I.~R.,
  {Sarajedini}, A., {Dotter}, A., {Chaboyer}, B., {Marin-Franch}, A.,
  {Majewski}, S., {Aparicio}, A., {Hempel}, M., {Paust}, N.~E.~Q., {Reid},
  I.~N., {Rosenberg}, A., \& {Siegel}, M. 2007, ApJ in press (arXiv:0709.3762)

\bibitem[{{Piotto} {et~al.}(2007){Piotto}, {Bedin}, {Anderson}, {King},
  {Cassisi}, {Milone}, {Villanova}, {Pietrinferni}, \& {Renzini}}]{piotto07}
{Piotto}, G., {Bedin}, L.~R., {Anderson}, J., {King}, I.~R., {Cassisi}, S.,
  {Milone}, A.~P., {Villanova}, S., {Pietrinferni}, A., \& {Renzini}, A. 2007,
  \apjl, 661, L53

\bibitem[{{Pritzl} {et~al.}(2005){Pritzl}, {Venn}, \& {Irwin}}]{pritzl05}
{Pritzl}, B.~J., {Venn}, K.~A., \& {Irwin}, M. 2005, \aj, 130, 2140

\bibitem[{{Siegel} {et~al.}(2007){Siegel}, {Dotter}, {Majewski}, {Sarajedini},
  {Chaboyer}, {Nidever}, {Anderson}, {Mar{\'{\i}}n-Franch}, {Rosenberg},
  {Bedin}, {Aparicio}, {King}, {Piotto}, \& {Reid}}]{siegel07}
{Siegel}, M.~H., {Dotter}, A., {Majewski}, S.~R., {Sarajedini}, A., {Chaboyer},
  B., {Nidever}, D.~L., {Anderson}, J., {Mar{\'{\i}}n-Franch}, A., {Rosenberg},
  A., {Bedin}, L.~R., {Aparicio}, A., {King}, I., {Piotto}, G., \& {Reid},
  I.~N. 2007, \apjl, 667, L57

\bibitem[{{Skrutskie} {et~al.}(2006){Skrutskie}, {Cutri}, {Stiening},
  {Weinberg}, {Schneider}, {Carpenter}, {Beichman}, {Capps}, {Chester},
  {Elias}, {Huchra}, {Liebert}, {Lonsdale}, {Monet}, {Price}, {Seitzer},
  {Jarrett}, {Kirkpatrick}, {Gizis}, {Howard}, {Evans}, {Fowler}, {Fullmer},
  {Hurt}, {Light}, {Kopan}, {Marsh}, {McCallon}, {Tam}, {Van Dyk}, \&
  {Wheelock}}]{2mass}
{Skrutskie}, M.~F., {Cutri}, R.~M., {Stiening}, R., {Weinberg}, M.~D.,
  {Schneider}, S., {Carpenter}, J.~M., {Beichman}, C., {Capps}, R., {Chester},
  T., {Elias}, J., {Huchra}, J., {Liebert}, J., {Lonsdale}, C., {Monet}, D.~G.,
  {Price}, S., {Seitzer}, P., {Jarrett}, T., {Kirkpatrick}, J.~D., {Gizis},
  J.~E., {Howard}, E., {Evans}, T., {Fowler}, J., {Fullmer}, L., {Hurt}, R.,
  {Light}, R., {Kopan}, E.~L., {Marsh}, K.~A., {McCallon}, H.~L., {Tam}, R.,
  {Van Dyk}, S., \& {Wheelock}, S. 2006, \aj, 131, 1163

\bibitem[{{Smith} {et~al.}(2000){Smith}, {Suntzeff}, {Cunha}, {Gallino},
  {Busso}, {Lambert}, \& {Straniero}}]{smith00}
{Smith}, V.~V., {Suntzeff}, N.~B., {Cunha}, K., {Gallino}, R., {Busso}, M.,
  {Lambert}, D.~L., \& {Straniero}, O. 2000, \aj, 119, 1239

\bibitem[{{Sneden}(1973)}]{moog}
{Sneden}, C. 1973, \apj, 184, 839

\bibitem[{{Stetson}(1981)}]{stetson81}
{Stetson}, P.~B. 1981, \aj, 86, 687

\bibitem[{{Sweigart}(1997)}]{sweigart97}
{Sweigart}, A.~V. 1997, \apjl, 474, L23

\bibitem[{{Villanova} {et~al.}(2007){Villanova}, {Piotto}, {King}, {Anderson},
  {Bedin}, {Gratton}, {Cassisi}, {Momany}, {Bellini}, {Cool}, {Recio-Blanco},
  \& {Renzini}}]{villanova07}
{Villanova}, S., {Piotto}, G., {King}, I.~R., {Anderson}, J., {Bedin}, L.~R.,
  {Gratton}, R.~G., {Cassisi}, S., {Momany}, Y., {Bellini}, A., {Cool}, A.~M.,
  {Recio-Blanco}, A., \& {Renzini}, A. 2007, \apj, 663, 296

\bibitem[{{Yong} {et~al.}(2006){Yong}, {Aoki}, {Lambert}, \&
  {Paulson}}]{yong06}
{Yong}, D., {Aoki}, W., {Lambert}, D.~L., \& {Paulson}, D.~B. 2006, \apj, 639,
  918

\bibitem[{{Yong} {et~al.}(2005){Yong}, {Grundahl}, {Nissen}, {Jensen}, \&
  {Lambert}}]{yong05}
{Yong}, D., {Grundahl}, F., {Nissen}, P.~E., {Jensen}, H.~R., \& {Lambert},
  D.~L. 2005, \aap, 438, 875

\end{thebibliography}
